# The Future of Computing is Boring
# (and that is exciting!)

How to get to Computing Nirvana in 20 years or less?

Aleksander Slominski
IBM T.J Watson Research Center
New York, NY
aslom@us.ibm.com

Vinod Muthusamy
IBM T.J Watson Research Center
New York, NY
vmuthus@us.ibm.com

Vatche Ishakian
*Computer Information Systems*
Bentley University, Waltham, MA
vishakian@bentley.edu

*Abstract*—We see a trend where computing becomes a metered utility similar to how the electric grid evolved. Initially electricity was generated locally but economies of scale (and standardization) made it more efficient and economical to have utility companies managing the electric grid. Similar developments can be seen in computing where scientific grids paved the way for commercial cloud computing offerings. However, in our opinion, that evolution is far from finished and in this paper we bring forward the remaining challenges and propose a vision for the future of computing. In particular we focus on diverging trends in the costs of computing and developer time, which suggests that future computing architectures will need to optimize for developer time.

## I. SHORT HISTORY OF UTILITY COMPUTING

The idea of treating computing as a utility similar to the electrical grid is not new. Early efforts starting in the 1960s were around time-sharing, mainframes and later datacenters to aggregate and share computing resources. The idea gained traction and became popular in 2000s. For example [GridBook] proposed "A computational grid is a hardware and software infrastructure that provides dependable, consistent, pervasive, and inexpensive access to high-end computational capabilities." Grids gained popularity in volunteer computing (for example Folding@home and Berkeley Open Infrastructure for Network Computing aka BOINC) to do very large distributed computations, for example Folding@home had 47 PFLOPS and more than 100K CPUs running [FoldingAtHomeStats].

From 2006 commercial vendors starting with Amazon Web Services (AWS) offer computing as an utility with Elastic Compute Cloud (EC2) bringing the "cloud computing" terminology to wide usage. Other vendors followed with their clouds such as Microsoft Azure, Google Cloud, IBM SmartCloud (later IBM Cloud), Oracle Cloud, and more.

The vision of the cloud can be described by 3 key characteristics that emphasize economical advantages: elasticity, low-barrier to start and pay-as-you-go [CloudVision]:

- Elasticity: the cloud provides "infinite" resources available on-demand;
- Low-barrier to start using cloud resources: only a credit card is needed and there is no long onboarding process;
- Pay-as-you-go: the pricing model allows users to avoid upfront costs and long term commitments.

Cloud computing emphasizes "on-demand" and "pay-as-you-go" aspects that were not the main priorities in previous distributed computing approaches. However the unit of computation was still a server even if it is packaged as virtual machines (VMs). That changed when containers gained popularity starting with Docker in 2013. Containers provide standardized packaging (into container images) and isolation for processes that are running in a containerized environment on top of one shared (virtual) machine instead of creating a VM for each computation (see the section below for more discussion about rise of containers). Kubernetes added orchestration for containers in 2014 to provide elasticity for container-based computation and cloud vendors provide it as a service with low-barrier to entry and pay-as-you-go pricing.

While we have still not completely realized the vision of utility computing on par with today's electrical grid, each new innovation, such as cloud computing, is bringing us closer to that goal.

## II. COST OF COMPUTING

Initially the cost of computing was very high and the cost of the machines far outweighed the cost of developer time - the computing machines were a large capital investment not easily shareable or available to rent. However, the cost of computing has decreased by orders of magnitude over time and the computing capabilities are now available in smaller units. There is no longer a need to buy whole servers but they can be rented. Conversely, as we will highlight below, the cost of developers is what tends to dominate computing today.

There is about 1500 to 2000 hours of business hours in one year (50 weeks x 40 hours more or less depending on vacations, holidays, etc.) and the cost of a developer ranges from $20 to $100 per hour on average resulting in the cost of a developer per year to be around $40K to $200K in Western countries. Costs in other parts of the world may be lower. We simplify for back-of-envelope calculations to show that the cost is high and is growing as demand for programmers keep growing outpacing supply. It is estimated there were around 20 million professional developers worldwide in 2017 [DevelopersWorldwide].

The average salary of a college graduate over last 60 years changed from $6K in 1960 to $50K in 2015, an 8x increase. However, if adjusted for inflation, the increase is only 6% (1.06x) [SalaryGraduates].

How does the cost of developers compare to the cost of computation over the years? We will look at how much computation could be bought with the amount of money equivalent to the cost of one developer - say, $100K. Let's look how much compute and memory we would get for that amount of money if we were buying a server machine over the last 60 years. Table I shows the cost for billion floating point operations per seconds (GFLOPS) [GflopsCost] and the cost of memory with one billion bytes (GB) [MemoryPrices] over last few decades.

TABLE I. COST OF COMPUTATION

| Year | Approximate cost per GFLOPS (2017 US dollars) | Approximate cost to buy GB of memory |
|---|---|---|
| 1960 | $150B* | $5B* |
| 1980 | $40M | $6M |
| 2000 | $1K | $840 |
| 2010 | $2** | $20 |
| 2020 (extrapolated from 2018) | $0.02** | $7 |

\* This is an extrapolated cost as there were hardware limitation on getting any machine with that level of performance.
\*\* Main advance in computation are by combining CPU with GPUs.

Table II shows how much computation (GFLOPS) we can buy with $100K. We also show the ratio of change: we can see that the rate of change slowed down for memory and GFLOPS with sustained improvements for CPUs only combining them with specialized computational units such as GPUs.

TABLE II. HOW MUCH COMPUTATION $100K BUYS

| Year | How many GFLOPs for $100K? (2017 US dollars) (ratio of change) | How many GBs of memory for $100K? (ratio of change) |
|---|---|---|
| 1960 | 0.0000006 GFLOPS or 600 FLOPS | 0.00002 GB or 20KB |
| 1980 | 0.0025 GFLOPS (3750x) | 0.016 GB (800x) |
| 2000 | 100 GFLOPS (40000x) | 119 GB (7437x) |
| 2010 | 50000 GFLOPS or 50 TFLOPS (500x) | 5000 GB or 5 TB (42x) |
| 2020 (extrapolated from 2018) | 5M GFLOPS or 5 PFLOPS (100x) | 15000 GB or 15 TB (3x) |

We can see several orders of magnitude more computational resources can be bought with the same amount of money over a 10 to 20-year span. However, that is not true about human resources as developer's time is getting more expensive when compared to how much compute time can be bought.

In our comparison, for the sake of simplicity, we have not included additional costs such as cost of physical location, maintenance, or electricity used both for running server and cooling. For example to have 5 PFLOPS requires to buy clusters of servers as currently CPU and GPU combination can provide about 100 TFLOPS. Similarly 10TB of memory requires to buy clusters of servers as the biggest servers today rarely can handle more than 6 TB of memory and common is to have about 64GB in one server. And for salary costs we did not include overhead costs such as office space, maintenance, or electricity for the office lights, heating, and cooling.

However, utility computing is no longer about buying servers. Instead it is about renting and paying only when using servers to run computations. Cloud VMs cost per GB of memory per hour is about $0.02 [RightScaleCompare] that is about 1000x to 5000x lower in cost than the developer's hourly rate of $20-$100 (for more details see the Table III below).

With VMs there are significant delays to add more capacity as it can only be expanded by starting new VMs (or deleting idle VMs). That was even worse before cloud computing as actual physical machines had to be ordered, delivered and installed. From having an order approved to a working machine could take months.

TABLE III. Evolution of computation

| Year | Computing Paradigm | Expenses | Time to solution | Cost of compute with 1 GB of memory per hour |
|---|---|---|---|---|
| 1960-2000 | Mainframes, data centers | Upfront capital investment | Months to Years | $100K to $1000 |
| 2000s | HPC Cluster and Grid Computing, data centers, VMs | Capital investment, reusing idle machines | Days to Weeks | $10-50 |
| 2010s | Cloud and Serverless Computing, VMs and Containers | Pay-as-you-go, sometimes paying even if not using compute resources (VMs) | Minutes to Hours | $0.01 - 0.06 |
| Future | Standardized containers | Pay only for what is used | Seconds to Minutes | <$0.01 |

An interesting recent development in cloud and utility computing is serverless computing that is providing smaller accounting time units where customers only pay for computation when it is running in contrast to VMs where accounting starts when the VM is booting and includes the time when nobody is using VMs. For serverless computing the unit of accounting is not hours or seconds but milliseconds. Therefore, for simplicity we use CPU hour/GB as the cost metrics, which is then about $0.06 in 2018 and we expect the cost to go down (see the Table III).

### III. THE RISE OF CONTAINERS

The key innovation allowing serverless computing is use of operating-system-level containers instead of VMs. VMs are a kind of container as they are packaged into a VM file format that allows to move and run them anyplace as long as there was VM runtime infrastructure available and the VM file format is compatible. That is however different from more light-weight containers that use operating-system-level virtualization. In this kind of containers instead of VMs providing virtualized hardware to run software, the operating system is managing containers that run software by accessing shared operating system (OS) that may be running on physical or virtualized hardware. When containers sharing the same OS then the resources such as memory can be much better used (density) and containers can be started much faster as they do not need to emulate startup process of physical machines that VMs need to do.

We will refer to operating-system-level containers simply as containers.

Even though eventually standardized format for VMs was created [OVF] still the inherent limitations of VMs in achieving high density and time wasted on startup makes them less attractive for utility computing where providers want containers that start fast and can be stopped fast too (so the time when container is running computation is maximized) and they want to package as many containers as possible in physical machines (density).

Containers gained recently popularity with Docker [Docker] and Open Container Initiative (OCI)[OCI] standardization of Linux containers provides well define file format to share containers.

Containers can now run on small IoT devices (such as Raspberry Pi) to large supercomputing clusters and they are gaining popularity as a leading way to run computations anywhere fulfilling dream that computational containers may have the same effect as intermodal shipping containers did on international trade after they were standardized even though it took decades to reach their full potential [ShippingContainers]

### IV. VISION

As we showed in previous sections, the cost of computation has dramatically lowered and the computation capabilities are becoming available on-demand as an utility. However, the cost of developer time has not decreased but is instead increasing (or not changing when accounting for inflation).

Our main thesis for future of computing is that the computing needs not only to be like an electrical utility, where it is always available and standardized, but it also needs to be optimized for developer productivity as it relates to business outcomes.

The computation in future need to work more like any other business where worldwide standardization drives cost down and makes it possible to transfer computing skills across domains. One promising development is the standardization of containers.

We expect that in the future business users and developers will use standardized containers to run business logic. Containers will be created and shipped to one or multiple cloud providers and be running on-demand in the cloud, edge, IoT or any other place where a container can be run.

Business users will have full flexibility in selecting providers and creating smart contracts that will automatically negotiate service level agreements (SLAs) with the best rates and move workloads between providers based on service level objectives (SLOs) and business key performance indicators

(KPIs). They will be able to track dependencies and set SLAs and SLOs for services used by business containers. New types of composition and orchestration services will make it easy to create new business applications built on containers and services. Blockchain could be used as a reliable ledger and run analytics across multiple customers with pseudo-anonymity, etc. By tracking the cost of computation and related KPIs (including revenue), business users will have detailed views of code and computing services, and their interplay. The developers and business users will see the relation between business outcomes (KPIs) and computing infrastructure (SLOs). Comprehensive continuous integration and delivery (CI/CD) pipelines will deploy new versions of business code with staged deployments based on SLOs and KPIs continuously evaluated for new and old versions of the code, or different deployment configurations. The new code will be deployed gradually in stages and automatically rolled back in case of problems. Business users can easily run A/B tests and react to results quickly.

Developers will have full visibility into all running containers and use tracing, logging, and monitoring infrastructure to find and automatically fix problems related to SLOs and KPIs. Machine Learning (ML) and Artificial Intelligence (AI) tools will be used to analyze both low level performance (SLOs) and their relation to metrics of business value (KPIs) and be allowed to deploy fixes to known problems. AI and ML may be used to analyze business applications for potential bottlenecks and then to deploy staged improvements or even deploy small long-tail applications and track how well they perform before deploying them to the next stage (see use cases below). Humans will work with AI and ML to validate and test new directions, new AI models may be deployed in containers and tested the same way the traditional software is.

## V. USE CASES - ONE DAY IN FUTURE OF COMPUTING

Let's consider a simple use case of introducing new a feature into a business application. Adam (a developer) is talking with Eva (a business user) and they wonder if a new insurance product is fulfilling customer expectations. They decide to run a small business trial, where they test a product with a small number of users. They modify the business logic to ask customers for feedback in specific situations. When code is committed to the version control system, the CI/CD process builds a new container that is immediately deployed in production after passing all unit and integration tests. The code is limited to a small set of customers selected by Adam and Eve, but can then be deployed in small increments to other customers. Adam and Eve monitor as the new code is working and can revert to an old version if needed.

Another use case is about ad-hoc application or seasonal computing, applications that are needed rarely and used heavily for short amounts of time with bursty workloads. For example during a natural disaster there may be need to develop and deploy a custom application to track and connect first responders and potential victims. With future computing it will become much easier to customize applications to particular circumstances and then deploy them at a large scale and only pay when they are used.

## VI. WHY BORING COMPUTING IS EXCITING

The Internet gained popularity and global interoperability by relying on a common TCP/IP stack. It hides the complexity of underlying network devices and providers and allows building applications on top of TCP/IP without the need to know how packets are processed by underlying layers below TCP/IP. This model is frequently visualized with an hourglass that has TCP/IP in the middle, with the network layer below and application layer on top. This model also supports the end-to-end principle [End2EndArg] where two applications implement more advanced application-specific functionality with the layer between them providing only necessary building blocks.

To fully realize the vision of future of computing we need an architecture similar to the Internet but for computing. We propose that future computing should be defined by an hourglass model with standardized containers in the middle, applications and programming models for composition and orchestration on top, and details of the container infrastructure at the bottom (cloud providers, IoT, etc.).

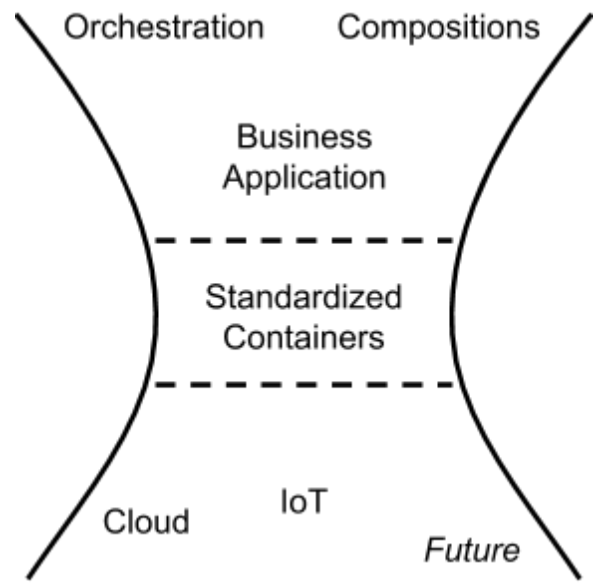

Fig. 1. Hourglass for future computing with containers serving as one shared standard

The container standardization should be rich enough to run the majority of today's applications. To support legacy computing code, it may be required to support running old computing non-standardized containerization formats (such as VMs) wrapped as a container inside a standardized container.

## VII. WHAT IS MISSING TODAY?

We have described the trends and current developments and many pieces are available.

The critical piece that we think is missing is to focus on improving human productivity as it has the greatest potential for improvements. In particular it should be much easier to make cost-based decisions for business users. Cloud services and serverless computing are examples of that direction. Improving developer's productivity should become top priority for future of computing research and software development. When it is fully acknowledged that developer time is the most precious resource then using it wisely becomes the most profitable industry - tools for computation management (code development, deployment, monitoring etc.) and for collaboration will provide the biggest economical gains. For example, investing into new technologies such as Augmented or Virtual Reality (AR/VR) may become very attractive if it is shown to help bring developers together and allow collaboration (such as virtual pair programming) that otherwise would be more expensive (removing travel time and other physical relocation costs).

Today cycles of development are long and business users have limited visibility into how computing is implementing business goals (KPIs) and how it translates to SLOs. Choosing SLAs is very time consuming and typically cloud providers are changed rarely and multi-cloud or hybrid-cloud support is very limited. In future contracts for using computing should be smarter and take advantage of markets and tracks SLAs.

Cloud computing and containers must be standardized (containers and how they are used) and become available everywhere where it makes economic sense. Legacy computing would be also containerized, perhaps as virtual machines (VMs) inside containers. From an economical point of view, it may be cheaper to keep legacy code running than pay for developers to re-design and re-write code using new computing approaches.

One possible bad news that may be long term good news for the future of computing is that Moore's law has ended by 2013 [MooresLaw]. That means the software can no longer depend on exponential progress in hardware performance. Instead software needs to get better with managing compute units that no longer keep changing and can be optimized for containers.

## VIII. CONCLUSION

Cloud computing as we know it today will no longer exist. It will evolve to become a ubiquitous computing layer that is everywhere, and architectural decisions will be driven only by cost vs value tradeoffs. That tradeoff will become clearly visible and easy to understand by business users that can depend on computing in the future the same way we depend on electricity today.

In the future computing must become like an electric utility for its users to get the benefits. What will we call this future computing infrastructure? Perhaps simply computing? Like electricity today, it is just there, and we no longer think about it unless it is not working.

When computing becomes boring, the future will be exciting. The exciting future can then be built on top of utility computing and provide necessary foundations to create solutions quickly that can be instantly deployed to make anywhere on Earth (and beyond). The future worth working for?